\documentclass[a4paper,12pt]{article}
\usepackage{amsmath,amsfonts,amssymb,amscd,enumerate}

\newtheorem{theorem}{Theorem}
\newtheorem{lemma}{Lemma}

\newtheorem{definition}{Definition}
\newtheorem{example}{Example}

\def\A{\mathcal{A}}
\def\C{\mathbb{C}}

\def\R{\mathbb{R}}
\def\Z{\mathbb{Z}}
\def\g{\mathfrak{g}}
\def\h{\mathfrak{h}}

\def\Ad{\text{Ad}}

\def\qed{$\blacksquare$}
\def\sech{\mbox{sech}}

\begin{document}

\title{On the Birkhoff factorization problem for the Heisenberg magnet and nonlinear
Schr\"{o}dinger equations}

\author{Sa\v{s}a Kre\v{s}i\'{c}-Juri\'{c}}

\date{}

\maketitle

\begin{abstract}
A geometrical description of the Heisenberg magnet (HM) equation with classical spins is given in terms of flows on the
quotient space $G/H_+$ where $G$ is an infinite dimensional Lie group and $H_+$ is a subgroup of $G$. It is shown that
the HM flows are induced by an action of $\R^2$ on $G/H_+$, and that the HM equation can be integrated by solving a
Birkhoff factorization problem for $G$. For the HM flows which are Laurent polynomials in the spectral variable we
derive an algebraic transformation between solutions of the nonlinear Schr\"{o}dinger (NLS) and Heisenberg magnet
equation. The Birkhoff factorization problem for $G$ is treated in terms of the geoemetry of the Segal-Wilson
Grassmannian $Gr(H)$. The solution of the problem is given in terms of a pair of Baker functions for special subspaces
in $Gr(H)$. The Baker functions are constructed explicitly for subspaces which yield multisoliton solutions of NLS and
HM equations.
\end{abstract}

\newpage

\section{Introduction}

From the work of Zakharov-Shabat \cite{ZS} and Sato-Segal-Wilson \cite{SW} it is known that completely integrable
partial differential equations, such as the Korteweg-de Vries (KdV) or the nonlinear Schr\"{o}dinger (NLS) equation,
are related with loop groups and infinite dimensional Grassmannians. The aim of this paper is to describe the
Heisenberg magnet (HM) equation
\begin{equation}\label{HM_equation}
\frac{\partial \vec S}{\partial t} = \vec S \times \frac{\partial^2 \vec S}{\partial x^2}, \quad S_1^2+S_2^2+S_3^2=1,
\end{equation}
in the context of loop groups, and to explore the construction of its solutions from the point of view that is close to
that of Segal-Wilson's work on KdV. The HM equation is a completely integrable system for which integrability was
proved in \cite{ZT} and the inverse scattering transform was developed in \cite{T}. Eq. \eqref{HM_equation} is the
isotropic case of the Landau-Lifshitz (LL) equation $\vec S_t = \vec S\times \vec S_{xx}+\vec S\times J\vec S$ when the
interaction constants are given by $J=diag(J_0,J_0,J_0)$. The LL model was studied by a number of authors. In \cite{Mi}
it was integrated by the Riemann factorization problem on a torus. Soliton solutions using the dressing procedure and
real algebraic-geometric solutions using theta functions were found in \cite{Bo1} and \cite{Bo2}. For a historical
account of the HM and LL equations see \cite{FT}. For our consideration of particular interest is the work by Carrey
et. al. \cite{Carrey} in which a spectral curve for the zero-curvature form of the LL hierarchy was introduced. The
authors show that the LL flows are induced by a group action on an infinite dimensional homogeneous space, and that
solutions of the LL hierarchy can be constructed by an analogue of Birkhoff factorization for elliptic curves. However,
the factorization problem is fairly difficult to solve explicitly which poses an obstacle in computing the flows. The
motivation for the present work stems from the fact that computation of the HM flows is reduced to solving a Birkhoff
factorization for a subgroup of $\Lambda GL(2,\C)$, $GL(2,\C)$-valued loops defined on the unit circle. This has the
consequence that the homogeneous space on which the HM flows are defined is closely related to the Segal-Wilson
Grassmannian. We study solutions of the HM equation in terms of the geometry of the Grassmannian, and explore its
relation to solutions of the focusing NLS equation. We show that solutions of the NLS equation can be expressed in
terms of the Baker functions for certain subspaces of the Grassmannian. By using the gauge transformation between the
NLS and HM equations one can associate these subspaces to solutions of the HM equation. We also exhibit subspaces which
yield multisoliton solutions of NLS and find an algebraic transformation which maps these solutions to multisoliton
solutions of HM.

The paper is organized as follows. In Sec. II we give a brief account of the group theoretic approach to integrable
systems. We descibe a general construction of partial differential equations which can be formulated as the
zero-curvature condition on the Lie algebra of a Banach-Lie group $G$. The solutions of such equations are represented
by flows induced by an action of $\R^n$ on an infinite dimensional homogeneous space. The flows can be integrated by
solving a Birkhoff factorization problem for $G$. Sec. III discusses the HM equation within the framework of Sec. II.
We define a loop group $G\subset \Lambda GL(2,\C)$ and show that the HM flows are induced by an action of $\R^n$ on the
homogeneous space $G/H_+$ where $H_+$ is a subgroup of ``positive'' loops in $G$. We also show that by choosing a
different subgroup $G_+\subset G$ the action induces the NLS flows on $G/G_+$ which are related with the HM flows by a
gauge transformation. This transformation is interpreted as a map between the quotient spaces $\Gamma \colon G/G_+\to
G/H_+$. We show that if the NLS flows are Laurent polynomials in the spectral variable $z\in S^1$, then $\Gamma$ is an
algebraic transformation between the NLS and HM solutions. In Sec. IV we describe a method for solving the Birkhoff
factorization problem for NLS by modelling the space $G/G_+$ as the Segal-Wilson Grassmannian $Gr(H)$ of the Hilbert
space $H=L^2(S^1,\C^2)$. We associate subspaces in $Gr(H)$ to solutions of NLS and show that the NLS flows can be
computed explicitly in terms of a pair of Baker functions for such subspaces. By modifiying the ideas from \cite{SW} we
construct subspaces which yield the multisoliton solutions of NLS. These solutions are then mapped to the multisoliton
solutions of HM by the transformation $\Gamma$.

\section{Group theoretic formulation of integrable\\ systems}

In this section we give a brief account of the group theoretic construction of integrable systems which admit the
zero-curvature representation. A more detailed discussion of the subject can be found, for example, in \cite{Haak}.
\begin{definition}
Let $G$ be a Banach Lie group. We say that $G$ admits a Birkhoff factorization denoted $(G,G_-,G_+)$
if $G$ contains closed subgroups $G_-$ and $G_+$ such that $G_-\cap G_+=\{e\}$ and the
product $G_-G_+$ is open in $G$.
\end{definition}
Let $\g$ be the Lie algebra of $G$ with the Lie bracket $[\cdot ,\cdot]$. The set $G_-G_+$ is open in $G$ if and only
if $\g$ splits into a direct sum of subalgebras $\g=\g_- \oplus \g_+$ where $\g_{\pm}$ is the Lie algebra of $G_{\pm}$.
The Birkhoff factorization is modelled to generalize the factorization of $GL(n,\C)$ into upper and lower triangular
matrices to infinite dimensions. Let $X_1,X_2, \ldots X_n$ be pairwise commuting elements of $\g_+$, $[X_i,X_j]=0$, and
consider a differentiable action $\R^n \times G\to G$ defined by
\begin{equation}\label{E1}
\mathbf{t}\ast g = \exp\left(\sum_{i=1}^n t_i\, X_i\right) g
\end{equation}
where $\mathbf{t}=(t_1,t_2,\ldots ,t_n)$. If $g\in G_-G_+$, then for $\mathbf{t}$ in a neighborhood of $0\in \R^n$ we
have $\mathbf{t}\ast g \in G_-G_+$ because $G_-G_+$ is open in $G$. Hence, $\mathbf{t}\ast g$ can be factored in a
unique way as
\begin{equation}\label{E2}
\mathbf{t}\ast g = g_-(\mathbf{t})\, g_+(\mathbf{t})
\end{equation}
where $g_{\pm}(\mathbf{t})\in G_{\pm}$. We say that the action \eqref{E1} induces the flow $g_{\pm}(\mathbf{t})$ on
$G_\pm$. The element $X_i\in \g_+$ is called the infinitesimal generator of the $t_i$ flow. Note that the action
\eqref{E1} descends to an action on the quotient space $G/G_+$ by $\mathbf{t}\ast (gG_+) = (\mathbf{t}\ast g)G_+$, thus inducing the
flow $g_-(\mathbf{t})G_+$ on $G/G_+$.

Next we show that the flow $g_-(\mathbf{t})$ represents solutions to a hierarchy of partial differential equations
(PDE) in zero-curvature form on the Lie algebra of $\g_+$. Let $p_{+}\colon \g \to \g_{+}$ denote the orthogonal
projection. Observe that Eqs. \eqref{E1} and \eqref{E2} imply
\begin{equation}\label{Ad1}
\Ad(g_-^{-1})\, X_i = g_-^{-1}\frac{\partial g_-}{\partial t_i}+\frac{\partial g_+}{\partial t_i}g_+^{-1}.
\end{equation}
By projecting Eq. \eqref{Ad1} onto $\g_+$ we obtain the following system of differential equations:
\begin{equation}\label{system}
\frac{\partial g_+}{\partial t_i} = M_i(\mathbf{t})\, g_+(\mathbf{t}), \quad \mbox{where}\quad M_i(\mathbf{t}) =
p_+\left(\Ad \left(g_-^{-1}(\mathbf{t})\right) X_i\right), \quad 1\leq i \leq n,
\end{equation}
Since $[X_i,X_j]=0$, the $t_i$ and $t_j$ flows commute so the compatibility condition $\partial_i \partial_j
g_+=\partial_j \partial_i g_+$ yields the zero-curvature equation \cite{ZS}
\begin{equation}\label{ZCC}
\frac{\partial M_i}{\partial t_j}-\frac{\partial M_j}{\partial t_i}+[M_i,M_j]=0, \quad i,j=1,2,\ldots ,n.
\end{equation}
Eq. \eqref{ZCC} represents a hierarchy of partial differential equations for the vector fields $M_i(\mathbf{t})$. In
concrete realizations of integrable systems $G$ is a Banach loop group, and Eq. \eqref{ZCC} is equivalent with a system
of differential equations for matrix elements $\{u(\mathbf{t})\}$ of $M_i$ and $M_j$. The zero-curvature equation is an
evolution equation for $u(x,t)$ where $x=t_1$ is the space variable and $t=t_k$, $k\geq 2$, is the time variable in the
$k^{th}$ equation of the hierarchy. Since $u(\mathbf{t})$ can be calculated explicitly from $g_-(\mathbf{t})$,
$u(\mathbf{t})$ is represented by the flow $g_-(\mathbf{t})G_+$ on the homogeneous space $G/G_+$. The group theoretic
approach to integrable systems can be used to study explicit solutions, symmetries and conservation laws in terms of
these flows. Note that the map $g\mapsto g_-(\mathbf{t})$ is invariant under the right multiplication of $g$ by an
element of $G_+$. Hence we may assume that $g=g_-(0)$, so $g$ encodes initial data for Eq. \eqref{ZCC}. Clearly,
\begin{equation*}
\exp\left(\sum_{i=1}^n \Delta t_i\, X_i\right) \left(g_-(\mathbf{t})G_+\right) = g_-(\mathbf{t}+\Delta \mathbf{t})G_+,
\end{equation*}
which means that the left multiplication of $g_-(\mathbf{t})G_+$ by $\exp(\Delta t_k\, X_k)$ pushes $u(\mathbf{t})$ in
the $t_k$ direction by the amount $\Delta t_k$. In this sense the Birkhoff factorization linearizes the equation for
$u(\mathbf{t})$, hence the map $g_-(\mathbf{t}) \mapsto u(\mathbf{t})$ can be viewed as an abstract version of the
inverse scattering transform for Eq. \eqref{ZCC}.

\section{Birkhoff factorization for the Heinsenberg\\ magnet equation}

In this section we discuss the Heisenberg magnet equation from the geometrical viewpoint presented in Sec. II. We
define a loop group $G$ and show that the flows corresponding to the HM and NLS equations
are induced by an action of $\R^n$ on the homogeneous spaces $G/H_+$ and $G/G_+$, where $H_+$ and $G_+$ are subgroups
of ``positive'' loops in $G$. Furthermore, we show that the gauge transformation between the NLS and HM equations can
be interpreted as a map between the quotient spaces on which the flows are defined. For loops which are Laurent
polynomials in the spectral parameter $z\in S^1$ this leads to an algebraic transformation between solutions of the NLS
and HM equations.

In order to provide $G$ with a Banach structure we start by introducing the Wiener algebra (see Dorfmeister
\cite{Dorfmeister})
\begin{equation*}
\A = \Big\{f\colon S^1\to \C\mid f(z)=\sum_{n=-\infty}^{\infty} c_n z^n,\; \sum_{n=-\infty}^{\infty}
|c_n|<\infty\Big\}.
\end{equation*}
This is a Banach algebra relative to the norm $\|f\|_1=\sum_{n=-\infty}^\infty |c_n|$. The algebra $\A$ consists of
continuous functions on $S^1$ which have an absolutely convergent Fourier series. Let $gl(n,\A)$ denote the Banach
algebra of matrices with elements in $\A$ equipped with the commutator bracket and the norm $\|g\|=\sum_{i,j}
\|g_{ij}\|_1$. In view of Wiener's lemma \cite{Rudin} the matrix $g(z)$ is invertible if and only if $det(g(z))\neq 0$
for all $z\in S^1$. Hence, the group of invertible elements $GL(n,\A)=\{g\in gl(n,\A)\mid det(g(z))\neq 0 \; \forall\;
z\in S^1\}$ is a Banach-Lie group as an open submanifold of $gl(n,\A)$. Let $\sigma$ be a continuous automorphism of
the algebra $gl(n,\A)$ and define the group
\begin{equation*}
G=\Big\{g\in GL(n,\A)\mid \sigma (g)=g\Big\}.
\end{equation*}
$G$ is a closed submanifold of $gl(n,\A)$, and hence a Banach-Lie group with Lie algebra
\begin{equation*}
\g = \Big\{g\in gl(n,\A)\mid \sigma (g)=g\Big\}.
\end{equation*}
We shall use the above construction of loop group $G$ in order to derive the HM and NLS flows. We remark that in some
cases it is more convenient to use an involution $\tau$ on $gl(n,\A)$, $\tau^2=id$, and to consider the subgroup
$\widetilde G =\{g\in GL(n,\A)\mid \tau (g) g = I\}$. This construction includes twisted loop groups related to systems such 
as the modified KdV equation \cite{AHP} and the Neumann oscillator \cite{Kresic}. For more examples of loop groups and 
integrable systems see \cite{Dorfmeister}.

In order to relate the HM equation with the loop group $G$ define the automorphism $\sigma \colon gl(2,\A)\to gl(2,\A)$
by
\begin{equation*}
\sigma(g(z))=\begin{pmatrix} 0 & 1 \\ -1 & 0 \end{pmatrix} \overline{g(\bar z)}\begin{pmatrix} 0 & -1 \\ 1 & 0 \end{pmatrix}.
\end{equation*}
Evaluating the condition $\sigma (g)=g$ we obtain
\begin{equation}\label{G}
G=\Bigg\{g\in GL(2,\A) \mid g(z)=\begin{pmatrix} a(z) & b(z) \\ -\overline{b(\bar z)} & \overline{a(\bar z)}\end{pmatrix} \Bigg\}.
\end{equation}
Consider the subgroups
\begin{equation*}
H_- =\Big\{ h\in G\mid h(z)=\sum_{n=0}^\infty A_n z^{-n}\Big\} \quad \mbox{and}\quad H_+ =\Big\{ h\in G\mid
h(z)=I+\sum_{n=1}^\infty B_n z^n \Big\}.
\end{equation*}
Clearly, $H_-$ and $H_+$ are closed subgroups of $G$ and $H_-\cap H_+=\{I\}$. Furthermore, the Lie algebras of $H_-$
and $H_+$ decompose the Lie algebra of $G$ into a direct sum $\g = \h_-\oplus \h_+$, hence the set $H_- H_+$ is open in
$G$. Thus $(G,H_-,H_+)$ is a Birkhoff factorization for $G$. We will frequently use the Pauli spin matrices
\begin{equation*}
\sigma_1=\left(\begin{matrix} 0 & 1 \\ 1 & 0 \end{matrix}\right), \quad
\sigma_2=\left(\begin{matrix} 0 & -i\\ i & 0 \end{matrix}\right), \quad \mbox{and}\quad \sigma_3=\left(\begin{matrix}
1 & 0 \\ 0 & -1\end{matrix}\right).
\end{equation*}
Now consider the pairwise commuting elements $X_k(z)=\sigma z^k \in \h_+$ where $\sigma = i\sigma_3$, and define a
differentiable action of $\R^n$ on $G$ by
\begin{equation}\label{HM_action}
\mathbf{t}\ast h = \exp\left(\sum_{k=1}^n t_k \sigma\, z^k\right)\, h.
\end{equation}
For any $h\in H_-H_+$ we have a unique factorization $\mathbf{t}\ast h = h_-(\mathbf{t}) h_+(\mathbf{t})$ for $h\in H_-
H_+$ when $\mathbf{t}$ is near $0\in \R^n$ because $H_-H_+$ is open in $G$. Let $p_+\colon \g\to \h_+$ denote the
orthogonal projection onto strictly positive powers of $z$, and consider the matrices
\begin{equation}\label{HM_fields}
M_k(\mathbf{t}) = p_+\left( h_-^{-1}(\mathbf{t})\, \sigma z^k\, h_-(\mathbf{t})\right), \quad 1\leq k\leq n.
\end{equation}
Observe that $M_k$ is a matrix polynomial of order $k$ in the parameter $z$. According to the general scheme outlined
in Sec. II the matrices \eqref{HM_fields} satisfy the zero-curvature condition \eqref{ZCC}. The system of equations
obtained in this way will be called the HM hierarchy. The following result shows that the first equation in the HM
hierarchy is Eq. \eqref{HM_equation}, and that solutions of Eq. \eqref{HM_equation} are obtained from the factorization
$(x,t)\ast h=h_-(x,t)h_+(x,t)$ for some $h\in H_-H_+$.
\begin{lemma}
Consider the Birkhoff factorization $(x,t)\ast h = h_-(x,t)h_+(x,t)$ for some initial data $h\in H_-H_+$, and let
$h_-(x,t)=\sum_{k=0}^\infty A_k(x,t)\, z^{-k}$. Then $S(x,t)=A_0^{-1}(x,t)\, \sigma_3\, A_0(x,t)$ is the matrix
representation of a solution of the HM equation \eqref{HM_equation}.
\end{lemma}

\noindent\textbf{Proof} Substituting $h_-(x,t)=\sum_{k=0}^\infty A_k(x,t) z^{-k}$ into Eq. \eqref{HM_fields} we obtain
the matrix polynomials
\begin{equation*}
M_1 = (A_0^{-1}\, \sigma\, A_0)z, \quad M_2 = (A_0^{-1}\, \sigma\, A_0)z^2+[A_0^{-1}\, \sigma\, A_0,A_0^{-1} A_1]z.
\end{equation*}
We show that
\begin{equation}\label{ZCC_xt}
\frac{M_1}{\partial t}-\frac{M_2}{\partial x}+[M_1,M_2]=0
\end{equation}
is the zero-curvature representation of Eq. \eqref{HM_equation}. Define the matrices $S=A_0^{-1}\sigma_3 A_0$ and
$P=A_0^{-1} A_1$. Since $A_0=\left(\begin{smallmatrix} a_0 & b_0
\\ -\overline{b_0} & \overline{a_0}\end{smallmatrix}\right)$, $S$ is a Hermitian matrix of the form
\begin{equation*}
S=\begin{pmatrix} S_3 & S_1-iS_2 \\ S_1+iS_2 & -S_3 \end{pmatrix} = \sum_{k=1}^3 S_k \sigma_k
\end{equation*}
for some real-valued functions $S_k(x,t)$. Moreover, we have $S^2=I$ which implies $S_1^2+S_2^2+S_3^2=1$. From this it follows
that Eq. \eqref{ZCC_xt} is equivalent with the system of equations for $S$ and $P$
\begin{align}
i\frac{\partial S}{\partial x}+[S,[S,P]] &=0, \label{S1} \\
\frac{\partial S}{\partial t}-\frac{\partial}{\partial x}[S,P] &=0.  \label{S2}
\end{align}
By substituting the identity $[S,[S,P]]=2S[P,S]$ into Eq. \eqref{S1} we obtain $[S,P]=(i/2) S S_x$. Then Eq. \eqref{S2}
yields
\begin{equation}\label{25}
\frac{\partial S}{\partial t} = \frac{i}{2}\left[\frac{\partial^2 S}{\partial x^2} S +\left(\frac{\partial S}{\partial x}\right)^2\right].
\end{equation}
Finally, we note that $2(S_x)^2 = - S_{xx} S - S S_{xx}$, thus Eq. \eqref{25} becomes
\begin{equation}\label{HM_matrix_form}
\frac{\partial S}{\partial t} = \frac{1}{4i}\left[S, \frac{\partial^2 S}{\partial x^2}\right].
\end{equation}
After dilating the time variable $t\mapsto t/2$ we conclude that Eq. \eqref{HM_matrix_form} is equivalent with Eq.
\eqref{HM_equation}. \qed \newline
Thus, solutions of the HM equation are represented by the flows $h_-(x,t)H_+$ on the homogeneous space
$G/H_+$.

Next we discuss the gauge transformation between the focusing NLS and HM equations in the context of Birkhoff
factorization for $G$. Recall that the gauge transformation $\gamma_g \colon \g\times \g \to \g \times \g$ by an
element $g(x,t)\in G$ is defined by
\begin{equation*}
\gamma_g (U,V)=\left(\Ad (g)U+\frac{\partial g}{\partial x} g^{-1}, \Ad (g)V + \frac{\partial g}{\partial t}
g^{-1}\right).
\end{equation*}
This transformation preserves the zero-curvature condition, and two systems of equations are said to be gauge
equivalent if their zero-curvature representations are related by a gauge transformation.

It is well known that Eq. \eqref{HM_equation} is gauge equivalent with the focusing NLS equation
\begin{equation}\label{NLS}
i\frac{\partial u}{\partial t}-\frac{1}{2}\frac{\partial^2 u}{\partial x^2}-4 u |u|^2 =0,
\end{equation}
where $u(x,t)$ is a complex valued function \cite{ZT}. We show that on the group level the gauge transformation
$NLS\mapsto HM$ can be interpreted as a map between the quotient spaces $\Gamma \colon G/G_+\to G/H_+$ where
$(G,G_-,G_+)$ is the Birkhoff factorization for $G$ defined by the subgroups
\begin{equation}\label{Gpm}
G_-=\Big\{ g\in G\mid g(z)=I+\sum_{n=1}^\infty A_n z^{-n}\Big\}, \quad G_+=\Big\{ g\in G\mid g(z)=\sum_{n=0}^\infty B_n
z^n\Big\}.
\end{equation}
Clearly, $\g=\g_- \oplus \g_+$ where $\g_{\pm}$ is the Lie algebra of $G_{\pm}$. One can think of $(G,H_-,H_+)$ and
$(G,G_-,G_+)$ as two factorizations which differ in the normalization conditions: $h_+(0)=I$ and $g_-(\infty)=I$.

The NLS equation can be written in zero-curvature form as follows. Suppose that the action of $\R^n$ on $G$ is given by
\eqref{HM_action}, and consider now the factorization $\mathbf{t}\ast g = g_-(\mathbf{t}) g_+(\mathbf{t})$ for some
$g\in G_-G_+$. Let $q_+ \colon \g \to \g_+$ denote the orthogonal projection onto non-negative powers of $z$. Then the
matrix polynomials defined by
\begin{equation}\label{NLS_fields}
\widehat M_k(\mathbf{t}) = q_+\left(g_-^{-1}(\mathbf{t})\, \sigma z^k\, g_-(\mathbf{t})\right), \quad k=1,2,\ldots, n,
\end{equation}
satisfy the system of equations \eqref{ZCC} which is called the NLS hierarchy. It is not difficult to see that Eq.
\eqref{NLS} is the first equation in the hierarchy. Denoting $g_-(x,t)=I+\sum_{n=1}^\infty A_n(x,t) z^{-n}$ and
evaluating Eq. \eqref{NLS_fields} we obtain
\begin{equation*}
\widehat M_1 = \sigma z +[\sigma, A_1], \quad \widehat M_2 = \sigma z^2 +[\sigma,A_1]z + [\sigma,A_2]- A_1[\sigma,A_1].
\end{equation*}
Since $A_n=\left(\begin{smallmatrix} a_n & b_n \\ -\overline{b}_n & \overline{a}_n \end{smallmatrix}\right)$, the
matrices $\widehat M_1$ and $\widehat M_2$ have the form
\begin{equation}\label{NLS_fields2}
\widehat M_1 = \sigma z+2\begin{pmatrix} 0 & ib_1 \\ -\overline{ib_1} & 0 \end{pmatrix}, \quad \widehat M_2 = \sigma
z^2+2\begin{pmatrix} 0 & ib_1 \\ -\overline{ib_1} & 0 \end{pmatrix} z +2\begin{pmatrix} -i|b_1|^2 & v \\ -\overline{v}
& i|b_1|^2\end{pmatrix},
\end{equation}
where $v=i(b_2-a_1 b_1)$. It is easily verified that the zero-curvature condition \eqref{ZCC} for $\widehat M_1$ and
$\widehat M_2$ is equivalent with Eq. \eqref{NLS} for $u=b_1$. The NLS equation can also be obtained by a reduction as
a special case of the AKNS hierarchy \cite{GM}.

We have seen that the loop group $G$ defined by Eq. \eqref{G} admits two factorizations: $(G,G_-G_+)$ and $(G,H_-H_+)$.
In fact, the sets $G_-G_+$ and $H_-H_+$ are equal so we may denote them by $K$. Since each $k\in K$ can be factored
uniquely as $k=g_- g_+ = h_- h_+$ we can define a map $\Gamma \colon K/G_+ \to K/H_+$ by $\Gamma (g_- G_+)=h_- H_+$.
The elements $h_{\pm}$ are related to $g_{\pm}$ simply by $h_- = g_- B_0$ and $h_+ = B_0^{-1} g_+$ where $B_0$ is the
zero-order Fourier coefficient of $g_+$. Note that if $k(x,t)=(x,t)\ast g$ is the flow in $G$ based at $k(0,0)=g\in
G_-$, then the cosets $k(x,t) G_+$ and $k(x,t) H_+$ represent the NLS and HM flows respectively. Thus, on the group
level $\Gamma$ maps the NLS solutions to HM solutions. Moreover, since $g_+=B_0 h_+$ and the vector fields $(M_1,M_2)$
satisfy Eq. \eqref{system}, we conclude that $(M_1,M_2)$ and $(\widehat M_1,\widehat M_2)$ are related by
$(M_1,M_2)=\gamma (\widehat M_1,\widehat M_2)$ where $\gamma$ is the gauge transformation by $B_0^{-1}$. Thus, we have

\begin{lemma}
Let $\widetilde K$ be the set of flows $\left\{(x,t)\ast g \mid g\in G_-\right\}$. Then the diagram
\begin{equation*}
\begin{CD}
\g_+\times \g_+ @>{\gamma}>> \h_+\times \h_+ \\
@A{\widehat \pi}AA                    @AA{\pi}A \\
\widetilde K/G_+ @>{\Gamma}>> \widetilde K/H_+
\end{CD}
\end{equation*}
is commutative, where the maps $\widehat \pi (g_-G_+)=(\widehat M_1,\widehat M_2)$ and $\pi (h_-H_+)=(M_1,M_2)$ are
defined in terms of Eqs. \eqref{NLS_fields} and \eqref{HM_fields} respectively, and $\gamma$ is the gauge
transformation by $B_0^{-1}$.
\end{lemma}

If $g_-(x,t)$ is a Laurent polynomial in the spectral parameter $z\in S^1$, then $\Gamma$ leads to a simple algebraic
transformation between the NLS and HM solutions. Suppose for the moment that $g_-(x,t)$ has a pole of order $N$ at
$z=0$, so that
\begin{equation}\label{NLSfactorization}
(x,t)\ast g = \left(I+\sum_{k=1}^N A_k(x,t) z^{-k}\right) \left(\sum_{k=0}^{\infty} B_k(x,t) z^k\right)
\end{equation}
where $g=I+\sum_{k=1}^N A_k(0,0) z^{-k}$. Comparing the coefficients with $z^{-N}$ on both sides of Eq.
\eqref{NLSfactorization} we conclude that $B_0(x,t)=A_N^{-1}(x,t) A_N(0,0)$. The solution of the HM equation is thus
given by
\begin{equation*}
\begin{split}
S(x,t) &= B_0^{-1}(x,t) \sigma_3 B_0(x,t) \\
&= A_N^{-1}(0,0) A_N(x,t) \sigma_3 A_N^{-1}(x,t) A_N(0,0).
\end{split}
\end{equation*}
Denote $A_N(x,t)=\left(\begin{smallmatrix} a & b
\\ -\bar b & \bar a \end{smallmatrix}\right)$ and let $a(0,0)=a_0$, $b(0,0)=b_0$. A straightforward computation shows
that the elements of $S$ can be expressed as
\begin{equation}\label{vecS}
\begin{split}
S_3 &= \frac{(|a|^2-|b|^2) (|a_0|^2-|b_0|^2) + 4 Re(ab\, \overline{a_0 b_0})}{(|a|^2-|b|^2)(|a_0|^2-|b_0|^2)},  \\
S_1+iS_2 &= \frac{2\left[(|a|^2-|b|^2)\, a_0 \bar b_0+ab\, {\bar b_0}^2-\overline{ab}\,
a_0^2\right]}{(|a|^2-|b|^2)(|a_0|^2-|b_0|^2)}.
\end{split}
\end{equation}
Hence, if $g_-(x,t)$ has a pole at $z=0$, then the transformation \eqref{vecS} completely determines the vector $\vec
S=(S_1,S_2,S_3)$ from the lowest order Fourier coefficient of the NLS flow $g_-(x,t)$. We will show in the next section that
factorization \eqref{NLSfactorization} leads to multisoliton solutions of NLS. In this case the transformation $\Gamma \colon \widetilde
K/G_+ \to \widetilde K/H_+$ maps multisoliton solutions of NLS to multisoliton solutions of HM. Finally, we remark that
the correct choice of the loop group is important for obtaining the desired classes of solutions. For example, the NLS
equation can also be derived from the group $G\cap SL(2,\A)$, but the subgroup $(G\cap SL(2,\A))_-$ no longer contains Laurent
polynomials, and hence no soliton solutions.

As an example of transformation \eqref{vecS} consider the initial data
\begin{equation*}
g(z)=I+\begin{pmatrix} 0 & -i\alpha \\ -i\alpha & 0 \end{pmatrix} z^{-1}.
\end{equation*}
The time evolution of $g(z)$ determined by Eq. \eqref{NLSfactorization} is given by
\begin{equation*}
g_-(x,t)=I+\begin{pmatrix} -i\alpha \tanh (2\alpha x) & -i\alpha e^{-i\alpha^2 t} \sech (2\alpha x) \\ -i\alpha
e^{i2\alpha^2 t} \sech (2\alpha x) & i\alpha \tanh (2\alpha x)\end{pmatrix} z^{-1}.
\end{equation*}
Then the transformation \eqref{vecS} yields
\begin{equation*}
\begin{split}
S_1(x,t) &= 2\cos (2\alpha^2 t) \tanh(2\alpha x) \sech (2\alpha x), \\
S_2(x,t) &= -2\sin (2\alpha^2 t) \tanh(2\alpha x) \sech (2\alpha x), \\
S_3(x,t) &=2\sech^2 (2\alpha x)-1.
\end{split}
\end{equation*}
This is a solution of the HM equation which represents the magnetization vector $\vec S$ of unit length which
rotates about the $z$-axis.

\section{Explicit solution of the Birkhoff factorization problem}

In this section we describe a geometrical solution of the Birkhoff factorization problem for the focusing NLS equation
in terms of an infinite dimensional Grassmannian $Gr(H)$. Our approach is based on the
ideas of Segal and Wilson \cite{SW} who obtained solutions of the KdV equation in terms of the Baker function for
special subspaces in $Gr(H)$. By a similar procedure we construct a pair of Baker functions which yield solutions of
the NLS equation. The geometrical approach to the NLS equation has been studied by several authors. Guil and Ma\~nas
\cite{GM} have used the Grassmannian model in the study of self-similar solutions of the AKNS hierarchy. As a
by-product they characterized points in the Segal-Wilson Grassmannian which correspond to Nakamura-Hirota rational
solutions of the non-focusing NLS equation \cite{NH}. In \cite{Previato} Previato obtained solutions of the focusing
and non-focusing NLS in terms of theta functions for the corresponding hyperelliptic curve. Although the Grassmannian
model for integrable systems has been well studied, the solution of the factorization problem for NLS given here does
not seem to appear in the literature.

In the following we give a brief account of the Segal-Wilson Grassmannian of the Hilbert space $H=L^2(S^1,\C^2)$. More
details can be found in \cite{SW} and \cite{PS}. Let $H$ be the Hilbert space of square integrable functions on $S^1$
with values in $\C^2$, $f(z)=\sum_{k\in \Z} a_k z^k$, $a_k\in \C^2$ and $|z|=1$. The space $H$ has a natural
decomposition $H=H_+\oplus H_-$ into closed subspaces $H_+=\{\sum_{k\geq 0} a_k z^k\}$ and $H_-=\{\sum_{k<0} a_k
z^k\}$. The Grassmannian $Gr(H)$ is the set of closed subspaces $W\subset H$ such that the orthogonal projection
$p_+\colon W\to H_+$ is a Fredholm operator and $p_-\colon W\to H_-$ is a Hilbert-Schmidt operator. It is not difficult
to see that $Gr(H)$ is a Hilbert manifold modelled on the space of Hilbert-Schmidt operators $\mathcal{C}_2(H_+,H_-)$.
A chart around $W\in Gr(H)$ is the set $\mathcal{U}_W=\{G(T)\mid T\in \mathcal{C}_2 (W,W^\perp)\}$ where $G(T)=\{x+Tx
\mid x\in W\}$ is the graph of $T$, together with the map $\mathcal{U}_W \to \mathcal{C}_2 (W,W^\perp)$ defined by
$G(T)\mapsto T$. Since $W$ and $W^\perp$ are both infinite dimensional, the Hilbert spaces $\mathcal{C}_2(W,W^\perp)$
and $\mathcal{C}_2(H_+,H_-)$ are isomorphic.

Recall that the index of a Fredholm operator $T$ is defined by $ind(T)=dim(ker(T))-dim(coker(T))$. If $W\in Gr(H)$,
then the index of $p_+\colon W\to H_+$ is called the virtual dimension of $W$, $v.dim(W)$. The Grassmannian is not
connected since the connected components are indexed by the integers $v.dim(W)$. Only the component $Gr_0 (H)=\{W\in
Gr(H)\mid v.dim (W)=0\}$ will play a role in applications to NLS. The set $Gr_0(H)\cap \mathcal{U}_{H_+}$ is sometimes
called the ``big cell'', and has the following important property.

\begin{lemma}
(i) $W\in Gr_0(H)\cap \mathcal{U}_{H_+}$ if and only if $p_+\colon W\to H_+$ is an isomorphism.\\
(ii) If $W\in Gr_0(H)$, then $W\in \mathcal{U}_{H_+}$ if and only if $W\cap H_-=\{0\}$.
\end{lemma}

\noindent\textbf{Proof} (i) Suppose that $W\in Gr_0(H)\cap \mathcal{U}_{H_+}$. Then $p_+\colon W\to H_+$ has index zero
and the subspace $W$ is of the form $W=\{x+Tx\mid x\in H_+\}$ for some Hilbert-Schmidt operator $T\colon H_+\to H_-$.
If $p_+(x+Tx)=0$ then clearly $x=0$, hence $p_+$ is injective. Furthermore, $ind(p_+)=0$ implies that
$dim(coker(p_+))=0$. Thus, $p_+$ is both injective and surjective, and hence an isomorphism.

Now suppose that $p_+\colon W\to H_+$ is an isomorphism. Then clearly $ind(p_+)=0$, and hence $W\in Gr_0(H)$. Since the
projection $p_-\colon W\to H_-$ is a Hilbert-Schmidt operator, so is the product $T=p_-\, p_+^{-1} \colon H_+\to H_-$.
We note that $W=\{x+Tx \mid x\in H_+\}$ which proves that $W\in \mathcal{U}_{H_+}$. Thus, $W\in Gr_0(H)\cap
\mathcal{U}_{H_+}$. Part (ii) is proved in a similar fashion. \qed

The full group $GL(H)$ of bounded invertible operators with bounded inverse does not act on $Gr(H)$ as it does not
preserve the properties of the projections $p_{\pm}\colon W\to H_{\pm}$. However, the restricted general linear group
$GL_{res}(H)$ acts on $Gr(H)$. $GL_{res}(H)$ is the subgroup of $GL(H)$ consisting of operators whose block form
$\left(\begin{smallmatrix} a & b \\ c & d \end{smallmatrix}\right)$ with respect to the decomposition $H=H_+\oplus H_-$
has off-diagonal terms Hilbert-Schmidt: $b\in \mathcal{C}_2(H_-,H_+)$ and $c\in \mathcal{C}_2(H_+,H_-)$. The diagonal
terms are then automatically Fredholm. The action of $GL_{res}(H)$ is transitive since the orbit through $H_+$ is
$Gr(H)$.

We shall be interestred in the group $\Gamma_+$ of holomorphic maps on the unit disk $g\colon D_0\to GL(2,\C)$, $D_0 =
\{|z|\leq 1\}$. The elements of $\Gamma_+$ can be viewed as multiplication operators on $H$. A computation involving
expansion of $g(z)$ into the Taylor series around $z=0$ shows that if $g(z)$ is holomorphic, then the corresponding 
multiplication operator $M_g\colon H_+\oplus H_- \to H_+\oplus H_-$ has the block form $M_g =\left(\begin{smallmatrix} a & b \\ 0 & d
\end{smallmatrix}\right)$ where $a$ and $d$ are invertible and $b$ is Hilbert-Schmidt.
Hence $M_g\in GL_{res}(H)$, and so $\Gamma_+$ acts on $Gr(H)$. In fact, since $a$ is invertible, $\Gamma_+$ acts on the
connected component $Gr_0(H)$. By a similar argument it can be shown that the group $\Gamma_-$ of based holomorphic
maps $g\colon D_{\infty}\to GL(2,\C)$, $D_{\infty}=\{|z|\geq 1\}$, where $g(\infty)=I$, also acts on $Gr_0(H)$. Due to
the analytical structure of the group $G$ introduced in Sec. III we have $G_{\pm}\subset \Gamma_{\pm}$ (see Eq.
\eqref{Gpm}). In particular, the loop $\exp\left(\sum_{k=1}^n t_k \sigma z^k\right)\in \Gamma_+$ acts on $Gr_0(H)$. For
any $W\in Gr_0(H)\cap \mathcal{U}_{H_+}$ define the subspace
\begin{equation*}
W(\mathbf{t})=\exp\left(\sum_{k=1}^n t_k \sigma z^k\right)W.
\end{equation*}
Clearly, $W(\mathbf{t})\in Gr_0(H)\cap \mathcal{U}_{H_+}$ when $\mathbf{t}$ is near $0\in \R^n$ because
$\mathcal{U}_{H_+}$ is open and $\Gamma_+$ acts on $Gr_0(H)$. Hence, by Lemma 3(i) the orthogonal projection $p_+
\colon W(\mathbf{t})\to H_+$ is an isomorphism.
\begin{definition}
Let $W\in Gr_0(H)\cap \mathcal{U}_{H_+}$ and consider the isomorphism $p_+ \colon W(\mathbf{t})\to H_+$. The Baker
functions for the subspace $W$ are the unique elements $\Psi_1 (\mathbf{t},z)$, \\ $\Psi_2 (\mathbf{t},z) \in W$ such
that
\begin{equation}\label{B_functions}
 \exp\left(\sum_{k=1}^n t_k \sigma z^k\right) \Psi_1 (\mathbf{t},z) = p_+^{-1}(e_1), \quad
\exp\left(\sum_{k=1}^n t_k \sigma z^k\right) \Psi_2 (\mathbf{t},z) = p_+^{-1}(e_2),
\end{equation}
where $e_1=\left(\begin{smallmatrix} 1 \\ 0 \end{smallmatrix}\right)$,
$e_2=\left(\begin{smallmatrix} 0 \\ 1 \end{smallmatrix}\right)$.
\end{definition}

In the following we show that the flows of the NLS hierarchy can be expressed in terms
of the Baker functions $\Psi_1$ and $\Psi_2$. For certain subspaces $W$ this yields the multisoliton
solutions of NLS. Let $g\in G_-$ and consider the subspace $gH_+\in
Gr(H)$. Suppose for the moment that $gH_+$ contains a subspace $W$ such that $W\in Gr_0(H)\cap \mathcal{U}_{H_+}$,
and let $\Psi_1$, $\Psi_2$ be the Baker functions for $W$. Existence of such subspaces will be shown shortly.
Since $W\subset gH_+$ we have $\Psi_1=g f_1$, $\Psi_2=g f_2$ for some $f_1,f_2\in H_+$, and
\begin{equation*}
\exp\left(\sum_{k=1}^n t_k \sigma z^k\right) g f_1 = p_+^{-1}(e_1), \quad
\exp\left(\sum_{k=1}^n t_k \sigma z^k\right) g f_2 = p_+^{-1}(e_2).
\end{equation*}
This can be combined into the matrix equation
\begin{equation*}
\exp\left(\sum_{k=1}^n t_k \sigma z^k\right) g\, \left[f_1\mid f_2\right] =
\left[p_+^{-1}(e_1)\mid p_+^{-1}(e_2)\right].
\end{equation*}
We note that the matrices involved here have the form
\begin{equation*}
[f_1\mid f_2]=\sum_{k=0}^\infty B_k z^k, \quad [p_+^{-1}(e_1)\mid p_+^{-1}(e_2)] = I+\sum_{k=1}^\infty A_k z^{-k}.
\end{equation*}
If the matrix $[f_1\mid f_2]$ is invertible, then by uniqueness of the Birkhoff factorization $(G,G_-,G_+)$ we have
\begin{equation}\label{NLS_factorization2}
\exp\left(\sum_{k=1}^n t_k \sigma z^k\right) g = g_-(\mathbf{t}) g_+(\mathbf{t}), \quad g_{\pm}\in G_{\pm},
\end{equation}
where $g_+(\mathbf{t})= [f_1\mid f_2]^{-1}$ and $g_-(\mathbf{t})=\left[p_+^{-1}(e_1)\mid p_+^{-1}(e_2)\right]$ which is
precisely the factorization problem for NLS. The importance of the last relation is that in view of Eq. \eqref{B_functions} 
the flow $g_-(\mathbf{t})$ can be expressed in terms of the Baker functions:
\begin{equation}\label{NLS_Baker}
g_-(\mathbf{t}) = \exp\left(\sum_{k=1}^n t_k \sigma z^k\right) \left[\Psi_1\mid \Psi_2\right].
\end{equation}
This result can be formulated as
\begin{theorem}
Let $g\in G_-$ and let $W$ be a subspace of $gH_+\in Gr(H)$ such that $W\in Gr_0(H)\cap \mathcal{U}_{H_+}$. Then the flow
$g_-(\mathbf{t})$ given by Eq. \eqref{NLS_Baker} is the unique solution of the NLS factorization problem \eqref{NLS_factorization2}
where $\Psi_1$ and $\Psi_2$ are the Baker functions for $W$.
\end{theorem}

\begin{example}
(One-soliton solution)
\end{example}
Perhaps the simplest interesting example of a subspace in $Gr_0(H)\cap \mathcal{U}_{H_+}$ is the one which yields the
one-soliton solution of NLS. Its construction resembles the one-soliton space for the KdV equation given in \cite{SW}.
Consider the points in the unit disk $0<|p_i|<1$, $i=1,2$, and the parameters $\lambda,\mu \in \C^{\times}$,
$\lambda\neq \mu$. Define $W_1$ to be the $L^2$-closure of the space of functions $f\colon S^1\to \C^2$,
$f=\left(\begin{smallmatrix} f_1 \\ f_2\end{smallmatrix}\right)$, where $f_1$ and $f_2$ are holomorphic in $D_0$ except
possibly for a simple pole at $z=0$, and which satisfy the condition
\begin{equation}\label{condition1}
f_1(p_1)=\lambda f_2(p_1), \quad f_1(p_2)=\mu f_2(p_2).
\end{equation}
It is straightforward to verify that $W_1\in Gr_0(H)$ and $W_1\cap H_-=\{0\}$. It follows from Lemma 3(ii) that $W_1\in
Gr_0(H)\cap \mathcal{U}_{H_+}$, so $p_+\colon W_1(\mathbf{t})\to H_+$ is an isomorphism. In order to obtain the Baker
functions for the NLS equation let us write $x=t_1$, $t=t_2$, and suppress $t_k$ for $k\geq 3$. Denote
\begin{equation*}
p_+^{-1}(e_1) = \begin{pmatrix} az^{-1}+1 \\ cz^{-1} \end{pmatrix}, \quad p_+^{-1}(e_2) = \begin{pmatrix} bz^{-1}
\\ dz^{-1}+1 \end{pmatrix}.
\end{equation*}
Then the Baker functions for $W_1$ have the form
\begin{align*}
\Psi_1(x,t,z) &= \exp\left(-x\sigma z - t\sigma z^2\right) p_+^{-1}(e_1) =
\begin{pmatrix} \left(1+a z^{-1}\right)\, e^{-ixz-itz^2} \\
c z^{-1}\, e^{ixz+itz^2}\end{pmatrix}, \\
\Psi_2(x,t,z) &= \exp\left(-x\sigma z - t\sigma z^2\right) p_+^{-1}(e_2) =
\begin{pmatrix} b z^{-1}\, e^{-ixz-itz^2} \\
\left(1+d z^{-1}\right)\, e^{ixz+itz^2}\end{pmatrix}.
\end{align*}
Since $\Psi_1$ and $\Psi_2$ satisfy the condition \eqref{condition1} the coefficients $a,b,c,d$ are given by
\begin{alignat}{2}
a &= \frac{p_1 \mu e^{i\theta_2}-p_2 \lambda e^{i\theta_1}}{\lambda e^{i\theta_1}-\mu e^{i\theta_2}}, & \quad
b &= -\frac{\lambda \mu (p_1-p_2)}{\lambda e^{-i\theta_2}-\mu e^{-i\theta_1}}  \label{ab} \\
c &= \frac{p_1-p_2}{\lambda e^{i\theta_1}-\mu e^{i\theta_2}}, & \quad d &= -\frac{p_1 \lambda
e^{-i\theta_2}-p_2 \mu e^{-i\theta_1}}{\lambda e^{-i\theta_2} - \mu e^{-i\theta_1}},
\end{alignat}
where $\theta_k = 2(xp_k+tp_k^2)$, $k=1,2$. In view of the relation
\begin{equation*}
g_-(x,t)=\exp\left(x\sigma z+t\sigma z^2\right)[\Psi_1\mid \Psi_2] = I+\begin{pmatrix} a & b \\ c & d \end{pmatrix}
z^{-1},
\end{equation*}
for the Baker functions to generate solutions of the NLS equation we must have $g_-(x,t)\in G_-$, i.e. the matrix coefficients must
satisfy $c=-\overline{b}$ and $d=\overline{a}$. These conditions are satisfied provided $p_1 = \overline{p}_2$ and
$\lambda \overline{\mu} = -1$. Let us write $p_1=\alpha+i\beta$ and $\lambda = e^{-2\beta x_0} e^{i2\varphi}$
for some $\varphi, x_0\in \R$. Then it follows from Eq. \eqref{ab} that $a$ and $b$ have the particularly simple form:
\begin{align*}
a(x,t) &= -\alpha+i\beta \mbox{th}\left[2\beta \left(x+\alpha t + x_0\right)\right], \\
b(x,t) &= i\beta \exp\left[i2\left(\alpha x + (\alpha^2-\beta^2) t + \varphi\right)\right]\,
\mbox{sech}\left[2\beta \left(x+\alpha t+x_0\right)\right].
\end{align*}
According to the discussion in Sec. III (see Eq. \eqref{NLS_fields2}) the coefficient $b(x,t)$ is a solution of Eq.
\eqref{NLS}, which is the well-known one-soliton solution of NLS.

\begin{example}
($n$-soliton solution)
\end{example}
The $n$-soliton solution is obtained by a simple generalization of the above construction. Consider $n$ points in
the unit disk $0<|p_i|<1$, $1\leq i \leq n$, and $n$ parameters $\lambda_1,\lambda_2,\ldots ,\lambda_n\in \C^\times$.
Let $W_n$ the $L^2$-closure of the space of functions $f\colon S^1\to \C^2$ where $f_1$ and $f_2$ are holomorphic in
$D_0$ except possibly for a pole of order $n$ at $z=0$, and which satisfy the $2n$ conditions
\begin{equation}\label{conditions2}
f_1(p_j) = \lambda_j f_2(p_j), \quad f_1(\bar p_j) = \mu_j f_2(\bar p_j), \quad j=1,2,\ldots ,n,
\end{equation}
where $\lambda_j \bar \mu_j=-1$. The Baker functions for $W_n$ are given by
\begin{align*}
\Psi_1(x,t,z) &= \begin{pmatrix} \left(1+\sum_{k=1}^n a_k z^{-k}\right)\, e^{-i(xz+tz^2)} \\
\left(\sum_{k=1}^n c_k z^{-k}\right)\, e^{i(xz+tz^2)} \end{pmatrix}, \\
\Psi_2(x,t,z) &= \begin{pmatrix} \left(\sum_{k=1}^n b_k z^{-k}\right)\, e^{-i(xz+tz^2)} \\
\left(1+\sum_{k=1}^n d_k z^{-k}\right)\, e^{i(xz+tz^2)}\end{pmatrix}.
\end{align*}
Conditions \eqref{conditions2} yield the following system of equations for $a_k,b_k,c_k,d_k$:
\begin{alignat}{2}\label{system2a}
\sum_{k=1}^n \left(\lambda_j\, e^{i\theta_j}\, \frac{1}{p_j^k}\, c_k - \frac{1}{p_j^k}\, a_k\right) &= 1, & \quad
\sum_{k=1}^n \left(\mu_j\, e^{i\bar \theta_j} \frac{1}{\bar p_j^k}\, c_k - \frac{1}{\bar p_j^k}\, a_k\right) &= 1, \\ \label{system2b}
\sum_{k=1}^n \left(\frac{1}{\lambda_j\, e^{i\theta_j}} \frac{1}{p_j^k}\, b_k - \frac{1}{p_j^k}\, d_k\right) &= 1, &
\quad \sum_{k=1}^n \left(\frac{1}{\mu_j\, e^{i\bar \theta_j}} \frac{1}{\bar p_j^k}\, b_k - \frac{1}{\bar p_j^k}\,
d_k\right) &= 1,
\end{alignat}
where $\theta_j = 2(xp_j+tp_j^2)$, $1\leq j \leq n$. The consistency condition $\lambda_j \bar \mu_j = -1$ ensures that $c_k=-\bar b_k$
and $d_k = \bar a_k$. The $n$-soliton solution of NLS is then given as the quotient of the determinants $b_1
=\Delta_1/\Delta$, where
\begin{equation*}
\Delta=\prod_{j=1}^n \frac{1}{\lambda_j \mu_j}
\begin{vmatrix} q_1 e^{-i\theta_1} & q_1^2 e^{-i\theta_1} & \cdots & q_1^n e^{-i\theta_1} & -\lambda_1 q_1 & -\lambda_1
q_1^2 & \cdots & -\lambda_1 q_1^n \\
\vdots & & & & & & & \vdots \\
q_n e^{-i\theta_n} & q_n^2 e^{-i\theta_n} & \cdots & q_n^n e^{-i\theta_n} & -\lambda_n q_n & -\lambda_n q_n^2 & \cdots
& -\lambda_n q_n^n \\
\bar q_1 e^{-i\bar \theta_1} & {\bar q_1}^2 e^{-i\bar \theta_1} & \cdots & {\bar q_1}^n e^{-i\bar \theta_1} & -\mu_1
\bar q_1 & -\mu_1 {\bar q_1}^2 & \cdots & -\mu_1 {\bar q_1}^n \\
\vdots & & & & & & & \vdots \\
\bar q_n e^{-i\bar \theta_n} & {\bar q_n}^2 e^{-i\theta_n} & \cdots & {\bar q_n}^n e^{-i\bar \theta_n} & - \mu_n \bar
q_n & -\mu {\bar q_n}^2 & \cdots & -\mu_n {\bar q_n}^n
\end{vmatrix},
\end{equation*}
$q_j=1/p_j$, and $\Delta_1$ is obtained by replacing the first column of $\Delta$ by the vector $(\lambda_1 \ldots
\lambda_n\; \mu_1 \ldots \mu_n)^T$. The explicit form of $b_1$ becomes fairly complicated as $n$ increases. To conclude
our discussion we remark that by solving system \eqref{system2a}-\eqref{system2b} for the lowest order coefficients $a_n$ and $b_n$, and
applying the transformation \eqref{vecS} to $a_n$ and $b_n$ we obtain the $n$-soliton solution of the HM equation. Hence, 
we can associate solutions of the HM equation to the subspaces $W_n\in Gr(H)$ via the mappings 
$W_n\mapsto (a_n,b_n)\mapsto (S_1,S_2,S_3)$.

\end{document}